\documentstyle[12pt,epsf,epsfig]{article}
\newfont{\thiplo}{msbm10 scaled\magstep 2}
\newfont{\gothic}{eufb10 scaled\magstep 2}
\newfont{\unc}{eurb10} 
\newskip\humongous \humongous=0pt plus 1000pt minus 1000pt
\def\caja{\mathsurround=0pt}\def\eqalign#1{\,\vcenter{\openup1\jot \caja
        \ialign{\strut \hfil$\displaystyle{##}$&$ 
        \displaystyle{{}##}$\hfil\crcr#1\crcr}}\,}
\newif\ifdtup

\def\eqright #1\cr{\noalign{\hfill$\displaystyle{{}#1}$}}
\def\eqleft #1\cr{\noalign{\noindent$\displaystyle{{}#1}$\hfill}}

\def\oldreffmt#1{\rlap{[#1]} \hbox to 2\parindent{}}

\def\figfmt#1{\rlap{Figure {#1}} \hbox to 1in{}}

\def\sectioneq{\def\theequation{\thesection.\arabic{equation}}{\let
\holdsection=\section\def\section{\setcounter{equation}{0}\holdsection}}}%

\newcounter{holdequation}

\def\begineq #1\endeq{$$ \refstepcounter{equation}\eqalign{#1}\eqno
	(\theequation) $$}
\def\contlimit{\,{\hbox{$\longrightarrow$}\kern-1.8em\lower1ex
\hbox{${\scriptstyle (a\rightarrow0)}$}}\,}
\def\centeron#1#2{{\setbox0=\hbox{#1}\setbox1=\hbox{#2}\ifdim
\wd1>\wd0\kern.5\wd1\kern-.5\wd0\fi
\copy0\kern-.5\wd0\kern-.5\wd1\copy1\ifdim\wd0>\wd1
\kern.5\wd0\kern-.5\wd1\fi}}
\def\centerover#1#2{\centeron{#1}{\setbox0=\hbox{#1}\setbox
1=\hbox{#2}\raise\ht0\hbox{\raise\dp1\hbox{\copy1}}}}
\def\centerunder#1#2{\centeron{#1}{\setbox0=\hbox{#1}\setbox
1=\hbox{#2}\lower\dp0\hbox{\lower\ht1\hbox{\copy1}}}}
\def\lsim{\;\centeron{\raise.35ex\hbox{$<$}}{\lower.65ex\hbox
{$\sim$}}\;}
\def\gsim{\;\centeron{\raise.35ex\hbox{$>$}}{\lower.65ex\hbox
{$\sim$}}\;}

\def\super#1{\ifmmode \hbox{\textsuper{#1}}\else\textsuper{#1}\fi}
\def\textsuper#1{\newcount\holdspacefactor\holdspacefactor=\spacefactor
$^{#1}$\spacefactor=\holdspacefactor}

\def\getcite#1,{\advance\citenumber by1
\def\getcitearg{#1}\def\lastarg{@}
\ifnum\citenumber=1
\ref{#1}\let\next=\getcite\else\ifx\getcitearg\lastarg\let\next=\relax
\else ,\ref{#1}\let\next=\getcite\fi\fi\next}

\def\pom{{\rm P\kern -0.53em\llap I\,}}
\def\spom{{\rm P\kern -0.36em\llap \small I\,}}
\def\sspom{{\rm P\kern -0.33em\llap \footnotesize I\,}}

\relax
\def\contlimit{\,{\hbox{$\longrightarrow$}\kern-1.8em\lower1ex
\hbox{${\scriptstyle (a\rightarrow0)}$}}\,}
\def\upon #1/#2 {{\textstyle{#1\over #2}}}
\relax
\renewcommand{\thefootnote}{\fnsymbol{footnote}}

\sectioneq

\def\til#1{\centeron{\hbox{$#1$}}{\lower 2ex\hbox{$\char'176$}}}
\def\tild#1{\centeron{\hbox{$\,#1$}}{\lower 2.5ex\hbox{$\char'176$}}}
\def\sumtil{\centeron{\hbox{$\displaystyle\sum$}}{\lower
-1.5ex\hbox{$\widetilde{\phantom{xx}}$}}}

\begin{document} 

\begin{titlepage} 

\rightline{\vbox{\halign{&#\hfil\cr
&\today\cr}}} 
\vspace{0.25in} 

\begin{center} 

The Potential Importance of Low Luminosity and High Energy at the LHC

\medskip

Alan R. White\footnote{arw@anl.gov }

\vskip 0.6cm

\centerline{Argonne National Laboratory}
\centerline{9700 South Cass, Il 60439, USA.}
\vspace{0.5cm}

\end{center}

\begin{abstract} 
Low luminosity runs at higher LHC energy could
provide definitive evidence for an electroweak scale sextet quark sector of QCD
that produces electroweak symmetry breaking and dark matter within the bound-state S-Matrix of QUD - a massless, weak coupling, 
infra-red fixed-point, SU(5) field theory that might underly and unify the full Standard Model.
\end{abstract} 

\vspace{0.5in}

\begin{center}

{\it Presented at the XLIIIrd International Symposium on Multiparticle Dynamics, Chicago, Il USA.}
\end{center}

\renewcommand{\thefootnote}{\arabic{footnote}} \end{titlepage}

\section{Introduction}

The ``nightmare scenario'' seems to be emerging at 
the LHC. A resonance has been discovered that looks a lot like a Standard Model Higgs boson, but no other new physics is seen! In particular, the much anticipated manifestation of supersymmetry has not happened. 
Theoretical inconsistency implies there
must be more, but there is no indication what to look for and there is much concern that the current theoretical framework for new physics searches may be seriously misdirected. 

QUD\footnote{Quantum Uno/Unification/Unique/Unitary/Underlying 
Dynamics}  is a massless, weak coupling,  infra-red fixed-point, SU(5) field theory
that I have discovered as having a massive 
bound-state S-Matrix that is generated by infra-red chirality transition anomalies and that, uniquely,  contains the unitary Critical Pomeron. Unbelievably (almost!), QUD might
also underly and unify the S-Matrix of the full Standard Model. Moreover, unitarity may require the presence of a Higgs-like boson resonance.

If the QUD S-Matrix is the origin\footnote{The Standard Model could be reproducing the ``Unique Unitary S-Matrix''.} of the Standard Model then, in addition to small neutrino masses, the only ``new physics'' is an electroweak scale, strongly interacting, sextet quark
sector of QCD that provides electroweak symmetry breaking and dark matter, but is hard to isolate at large $p_{\perp}.$
The dynamicis is conceptually radical (within today's theory paradigm), and a calculational procedure has still to be developed. Nevertheless, the implications of QUD's existence are overwhelming. Suggestive evidence already exists and low luminosity runs at higher LHC 
energy could be definitive.



\section{The Critical Pomeron and the Formulation of QUD}

The Reggeon Field Theory Critical Pomeron is the only known description of rising total cross-sections\footnote{A necessity to match with an
asymptotically-free gauge theory at short distances.} that  
satisfies full multiparticle t-channel  unitarity and all s-channel
unitarity constraints. The supercritical phase occurs {\it \{uniquely\}} in superconducting QCD, and 
the critical behavior appears when asymptotic freedom is saturated. 
Saturation is achieved with 6 color triplet quarks and 2 color sextet quarks and is physically realistic
if  ``sextet pions'' produce electroweak symmetry breaking!

QCD sextet quarks with the right electroweak quantum numbers, plus the electroweak interaction, embed
uniquely (with asymptotic freedom and no anomaly) in 
QUD, i.e. SU(5) gauge theory with
left-handed massless fermions in the $5 \oplus 15 \oplus
40 \oplus 45^*$ representation.
Under $ SU(3)\otimes SU(2)\otimes
U(1)$
{\footnotesize 
$$
5=(3,1,-\frac{1}{3}))
+(1,2,\frac{1}{2})~,~~~~~ 15=(1,3,1)+
(3,2,\frac{1}{6}) + (6,1,-\frac{2}{3})~,
$$
$$
40=(1,2,-\frac{3}{2})
+(3,2,\frac{1}{6})+
(3^*,1,-\frac{2}{3})+(3^*,3,-\frac{2}{3}) + 
(6^*,2,\frac{1}{6})+(8,1,1)~,
$$
$$
45^*=(1,2,-\frac{1}{2})+(3^*,1,\frac{1}{3})
+(3^*,3,\frac{1}{3})+(3,1,-\frac{4}{3})+(3,2,\frac{7}{6}))+
 (6,1,\frac{1}{3}) +(8,2,-\frac{1}{2})
 $$}
Astonishingly, 
there are 3 ``generations'' of both leptons and triplet quarks, and QUD is vector-like wrt SU(3)xU(1)$_{em}$. SU(2)xU(1) is
not quite right, but if the anomaly-dominated
S-Matrix 
can be constructed via multi-regge theory, as I have outlined\cite{arw},
all elementary fermions are confined
and Standard Model interactions and states emerge.

\section{QUD Multi-Regge Theory}

In multi-regge limits, infinite momentum bound-states and interactions 
can be studied using $k{\scriptstyle \perp}$ reggeon
diagrams. The removal of fermion masses introduces
``anomaly vertices'' and after the (crucially ordered) removal of gauge boson masses and
a cut-off $k_{\perp}^{\lambda}$, an overall divergence\footnote{After elaborate cancelations of reggeization 
infra-red divergences.} produces a ``wee parton vacuum'' of universal anomalous wee 
gauge bosons (SU(5) adjoint ${\scriptstyle C\neq \tau}$) as illustrated in Fig.~1.
The surviving interactions couple via anomalies and  preserve the vector
SU(3)xU(1) symmetry. They are
\begin{enumerate}
\openup-1\jot {\item{\it Even Signature \{Critical\} Pomeron $\approx$ SU(3) gluon reggeon
+ wee gauge bosons. No
BFKL pomeron and no odderon.}
\item{\it Odd Signature Photon $\approx$ U(1)$_{em}$
gauge boson + wee gauge bosons.}
\item{\it Electroweak Interaction $\approx$ left-handed gauge  boson, 
mixed  with sextet pion (via anomalies), + wee gauge bosons.}}
\end{enumerate}
Anomaly color factors, in wee gauge boson infinite sums, enhance
couplings - hopefully to Standard Model values \{$\alpha_{\scriptscriptstyle QCD} ~>>~ \alpha_{\scriptscriptstyle QUD} \sim
\frac{1}{120}$\}
\begin{figure}
\centering
\includegraphics[width=0.75\columnwidth]{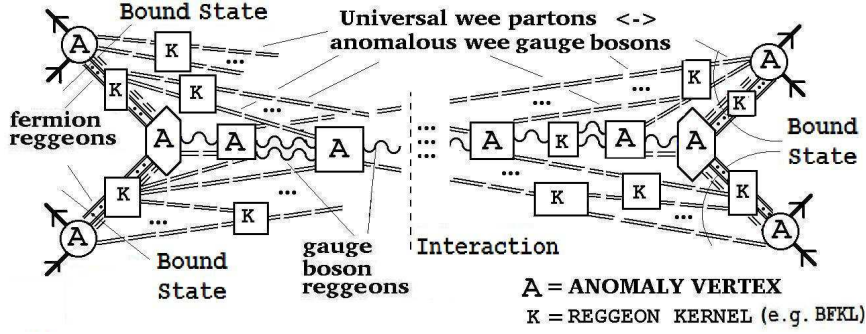}
\caption{The General Form of Divergent Reggeon Diagrams. \label{fig1}}
\end{figure}

Bound-states involve anomaly poles due to chirality transitions, e.g. Goldstone pions in QCD as illustrated in Fig.~2.  
\begin{figure}
\centering
\includegraphics[width=0.75\columnwidth]{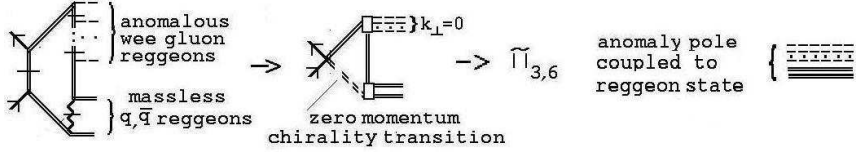}
\caption{The Anomaly Pole Goldstone Pion. \label(fig2)} 
\end{figure}
Within QCD, confinement and chiral symmetry breaking coexist with a ``parton model'' 
\begin{itemize}
\openup-1\jot{
\item{\it Bound-states are triplet or sextet quark mesons and baryons. The proton and 
neuson (``dark matter'' sextet neutron) are stable. There are no hybrids and no glueballs.}
\item{\it Sextet anomaly color factors are much larger than triplets and so (electroweak scale) sextet masses are correspondingly
larger.} 
\item{\it Wee gluon color factors
give large pomeron couplings to sextet states, producing
large high-energy x-sections and large  couplings to (pomeron producing) high-multiplicity hadron states.}}
\end{itemize}

Within QUD, ultra-violet octet
quark anomaly poles produce Standard Model generations of physical hadrons and leptons. Lepton bound states
contain three elementary leptons -{
\begin{itemize}\openup-1\jot{
\item{\it $(e^-,\nu)$ ~${\scriptstyle \leftrightarrow ~
(1,2,-\frac{1}{2}) \times \{(1,2,-\frac{1}{2}) (1,2,\frac{1}{2})\}_{AP} \times
\{(8,1,1)(8,2,-\frac{1}{2})\}_{UV}}$ }
\item{\it $(\mu^-,\nu)$ ~ ${\scriptstyle \leftrightarrow ~
(1,2,\frac{1}{2})\times  \{(1,2,-\frac{1}{2})  (1,2,-\frac{1}{2})\}_{AP} 
\times \{(8,1,1)(8,2,-\frac{1}{2})\}_{UV}}$}
\item{\it $(\tau^-,\nu)$~  ${\scriptstyle \leftrightarrow~
(1,2,-\frac{3}{2}) \times \{(1,2,\frac{1}{2})  (1,2,\frac{1}{2})\}_{AP} 
\times \{(8,1,1)(8,2,-\frac{1}{2})\}_{UV}}$}}
\end{itemize}
Anomaly interactions imply $M_{\scriptstyle hadrons} >> M_{\scriptstyle leptons} >>  M_{\scriptstyle \nu 's} ~\sim~
\alpha_{\scriptstyle QUD}$. 
The electron is almost elementary (the anomaly pole is, effectively, a minimal disturbance of the Dirac sea). The muon has the same constituents in a more massive dynamical configuration!
 
\section{``Top'' and ``Higgs'' Physics}

The primary decay of the ``sextet $\eta$'' 
is $\eta_6 \to W^+~W^-Z^0 ~\to~ W^+~W^-~b\bar{b}$ (cf. ${\scriptstyle \eta \to   \pi^+ \pi^- \pi^0}$). Since this is the
dominant Standard Model $t\bar{t}$ decay mode, the ${\eta_6}$ resonance produces
events that are experimentally hard to distinguish from Standard Model top physics. However, it is the sextet quark mass scale 
that is involved, and not a ``bizarrely large'' triplet quark mass!

Two QUD triplet quark generations give Standard Model hadrons. The third is 
$(3,-\frac{1}{3}) ~\equiv ~[3,1,-\frac{1}{3}] ~\in 5, ~~~~
(3,\frac{2}{3}) ~\in [3,2,\frac{7}{6}] ~\in 45^*$. The physical b quark is a mixture of all three QUD generations.
However, there are two ``exotic'' triplet quarks with charges -4/3 \& 5/3 that have no
chiral symmetry and so do not produce light (anomaly pole) bound-states. The left-handed ``top quark'' ($t_{QL}$) forms an 
electroweak doublet with one of the
exotic quarks and so will not appear in low mass states. (The electroweak interaction is enhanced, at high energy, by the sextet
quark sector.) The $\eta_t \approx t_{QR}\bar{t}_{QL}$ remains as a
``constituent $t_Q$ state". Mixing with the sextet $\eta$ gives two mixed-parity scalars
- the $\eta_6$ with electroweak scale mass and the $\eta_3$, with a mass 
between  triplet and sextet scales that, could  be $\sim$ 125 GeV! If so, the ``QUD Higgs'' is, predominantly a ``top/anti-top'' resonance.

Because QUD interactions are reggeized, intermediate state cancelations
must occur that are equivalent to the ``tree-unitarity'' condition determining electroweak Higgs couplings in the Standard Model. 
Consequently, the combined     
$\hat{\eta}_3$ and $\hat{\eta}_6$ couplings should reproduce Standard Model couplings and 
disparities between the 125 GeV resonance and the Standard Model Higgs should be accounted for by the 
$\eta_6$.
  
\section{At the LHC ?}
 
The most direct evidence for QUD is the appearance of the $\eta_6$ resonance, in the Z-pair x-section, at
the ``$t\bar{t}$'' threshold. As can be seen from Fig.~3, it is most visible in the lower luminosity 7 GeV data.
Is ultra-high luminosity missing QUD x-sections (as we will suggest often in the following)?
\begin{figure}
\centering
\includegraphics[width=0.35\columnwidth]{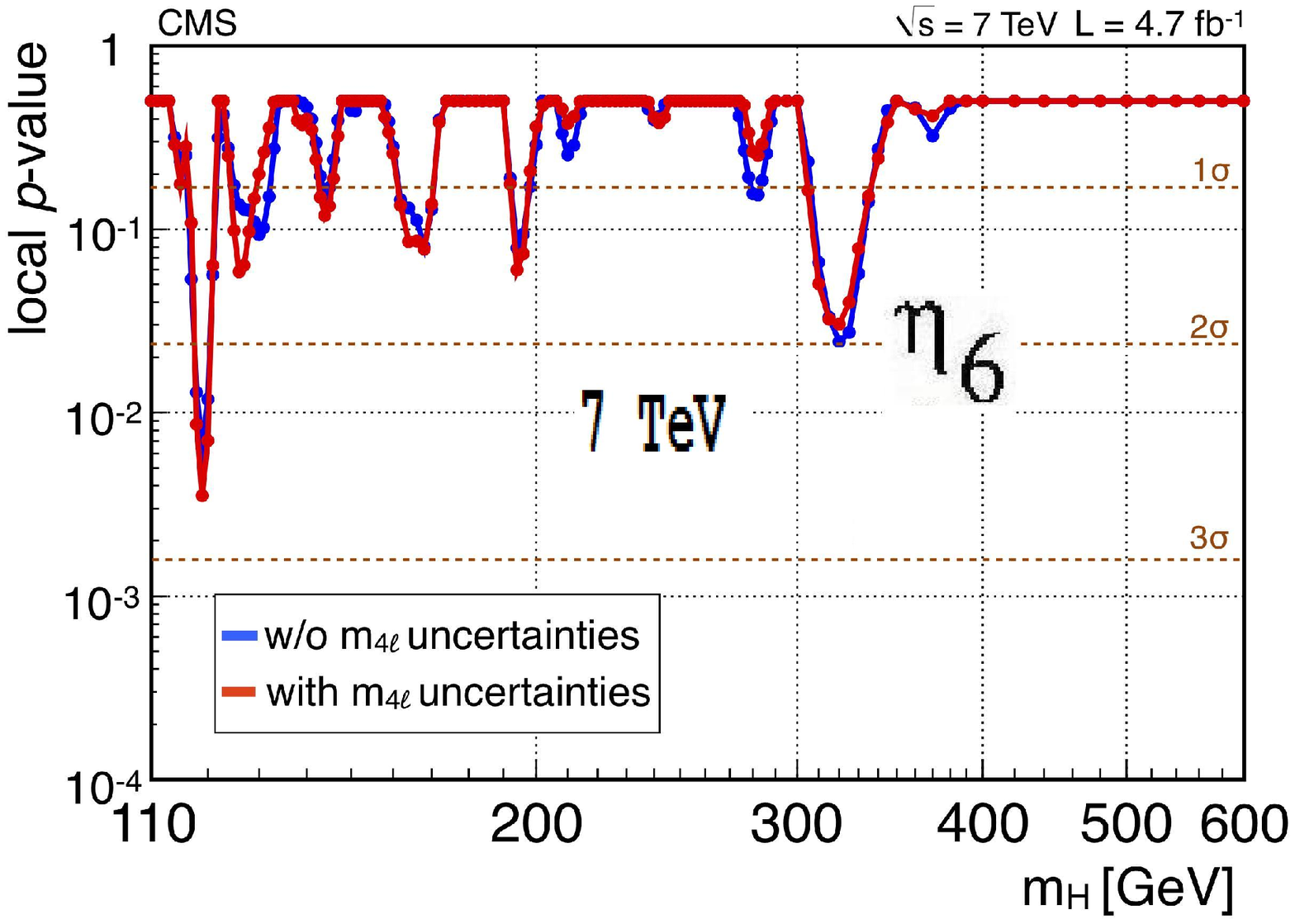} 
\includegraphics[width=0.35\columnwidth]{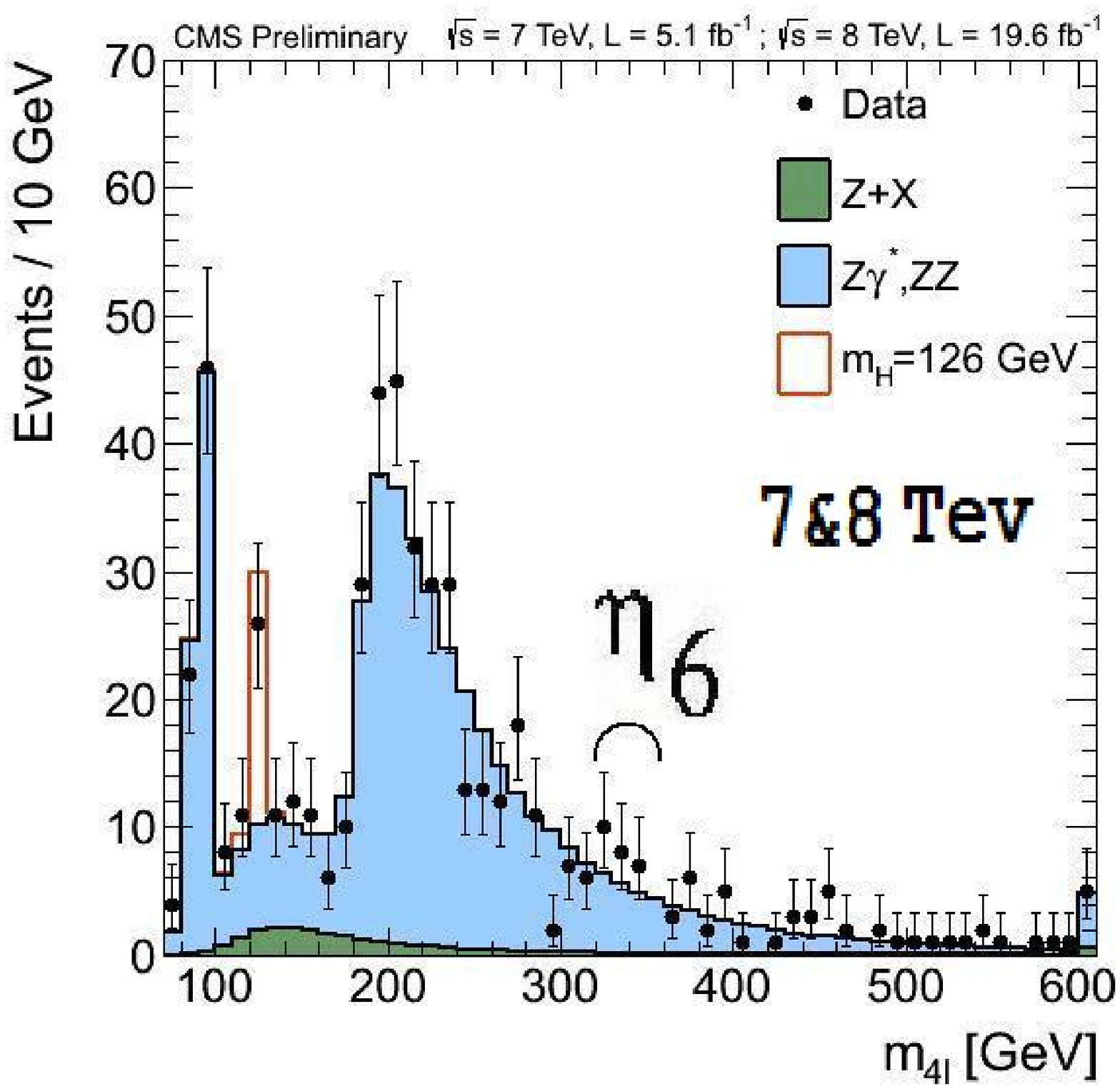}
\caption{The $\eta_6$ at the LHC. \label{fig3}}
\end{figure}

Large pomeron couplings to $\pi_6$'s 
\{$\equiv$ longitudinal W's/Z's\}
implies large rapidity-gap x-sections for multiple W's/Z's (i.e. sextet isospin conserving WW, ZZ, WWZ, ZZZ, ...) above the EW scale, including a large double-pomeron x-section for $Z^0Z^0$ and $W^+W^-$ pairs (some of these events might be identified, partially or fully, via jets). There should also be (correlated) much larger, x-sections for multiple W's/Z's, over a wide range of rapidities, with high associated hadron multiplicity - that are the intermediate states of the pomeron.
At higher energies, multiple sextet baryons
- ``neusons'' \{dark matter\} and ``prosons'' will be similarly produced. 
Growing x-sections, coupling pomeron and electroweak physics are clearly what should be looked for at the highest LHC
energy. However, {\it low luminosity is essential !!}

\noindent Existing evidence, appropriately interpreted, includes 
  \begin{enumerate} \openup-1\jot{
\item{\it ``Heavy Ion'' UHE cosmic rays are dark matter neusons.} 
\item{\it The spectrum knee is due to arriving/produced neuson
thresholds.} 
\item{\it Enhancement of high multiplicities and small $ p_{\perp}$
at the LHC reflects a sextet anomaly generated triple pomeron coupling.}
\item{\it ``Top quark events" are due to the $\eta_6$ 
resonance - interference with the background produces the Tevatron asymmetry.}
\item{\it $W^+W^-$ and $Z^0Z^0$ pairs have high mass excess x-sections, with the
 $\eta_6$ resonance appearing at the ``$t\bar{t}$ threshold".}
\item{\it The 125 GeV Higgs is the QUD \{$t_R\bar{t}_L $ + $\eta_6$\}
resonance.}
\item{\it The AMS $e^+/e^-$ ratio reflects EW scale CR production of W's \&
Z's  (+ neuson/antineuson annihilation?)}
\item{\it Low luminosity Tevatron/LHC events with a $Z$ pair + high
multiplicity 
of small  $p_{\perp}$ particles, could be QUD  (not Standard Model) events. }
\item{\it TOTEM+CMS missing momentum events could be $~ZZ ~\to \nu$'s }}
\end{enumerate}

\section{The Low Luminosity Events} 

Interesting events were seen with initial low luminosity
at the Tevatron and the LHC, and also in the
recent TOTEM-CMS low luminosity run. 
\begin{figure}
\centering
\includegraphics[width=0.3\columnwidth]{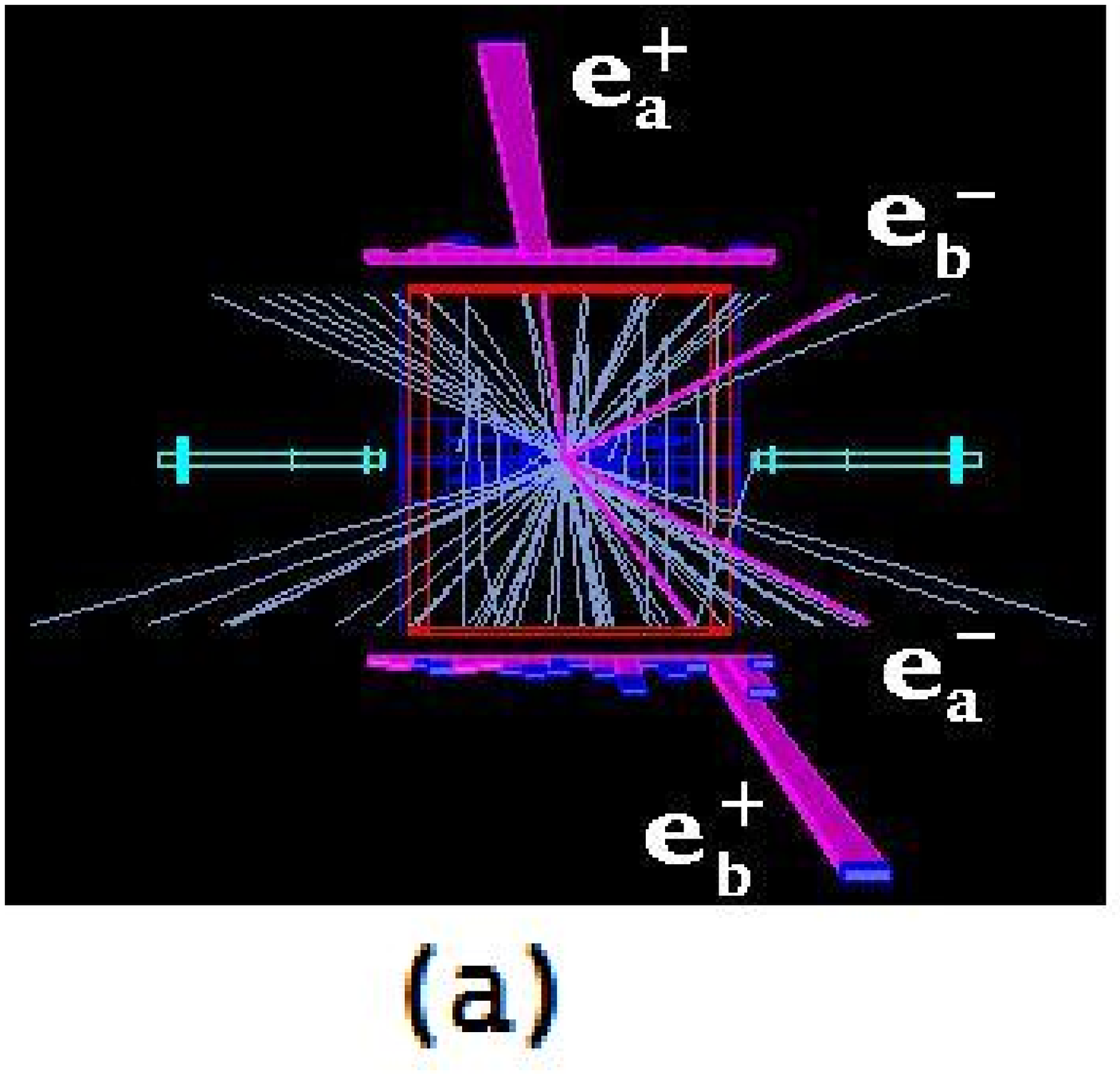} 
\includegraphics[width=0.4\columnwidth]{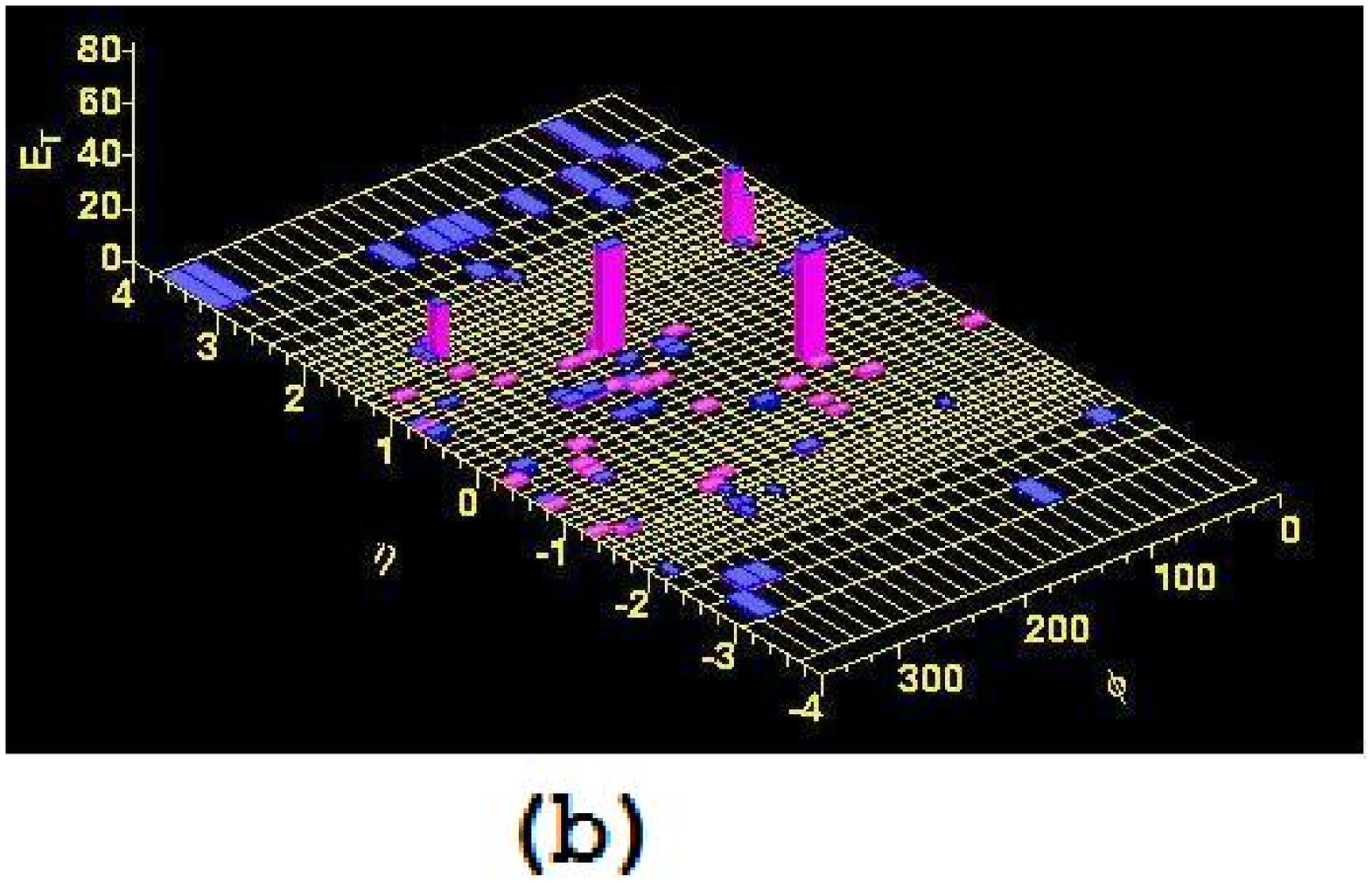}

\includegraphics[width=0.45\columnwidth]{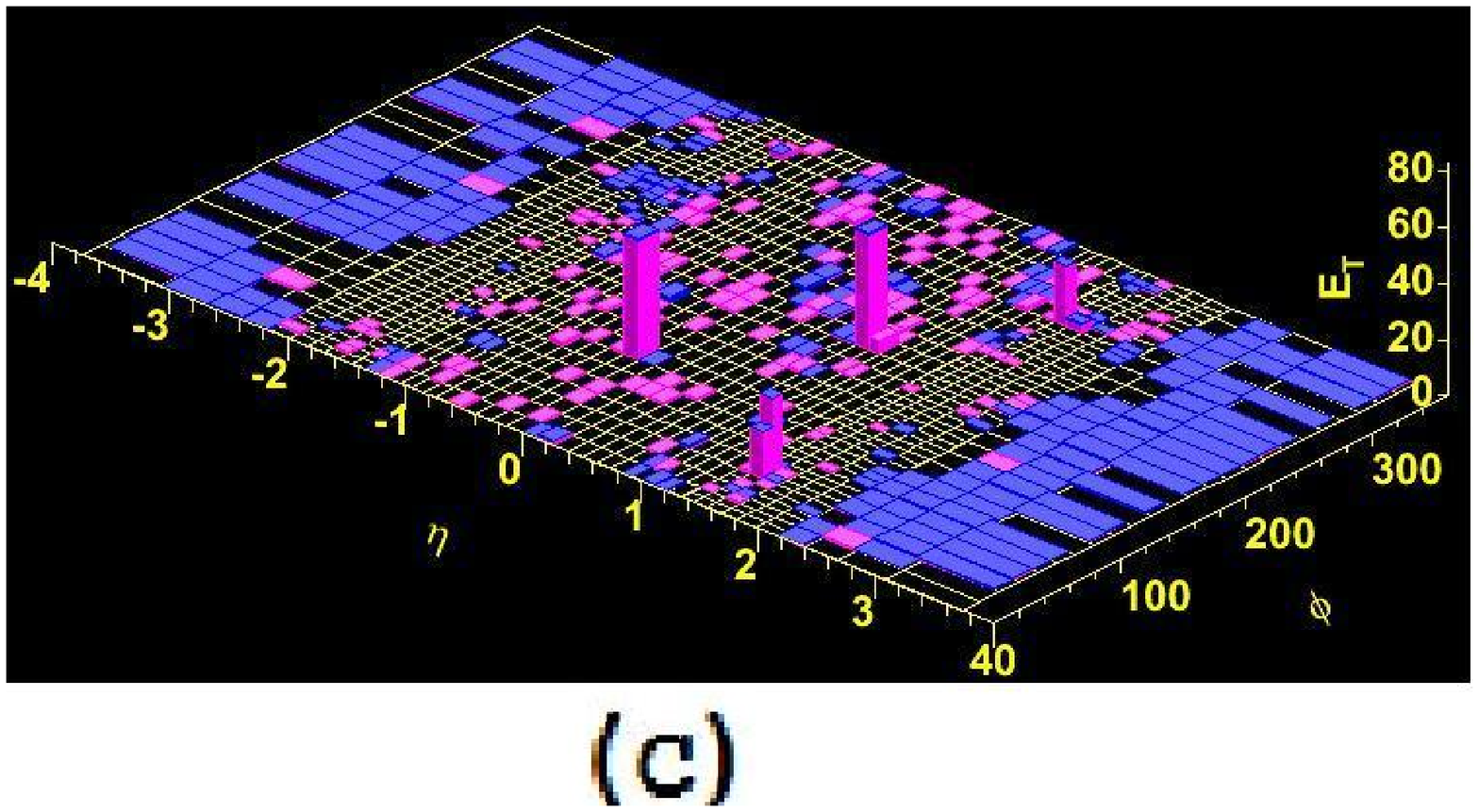} 
\includegraphics[width=0.45\columnwidth]{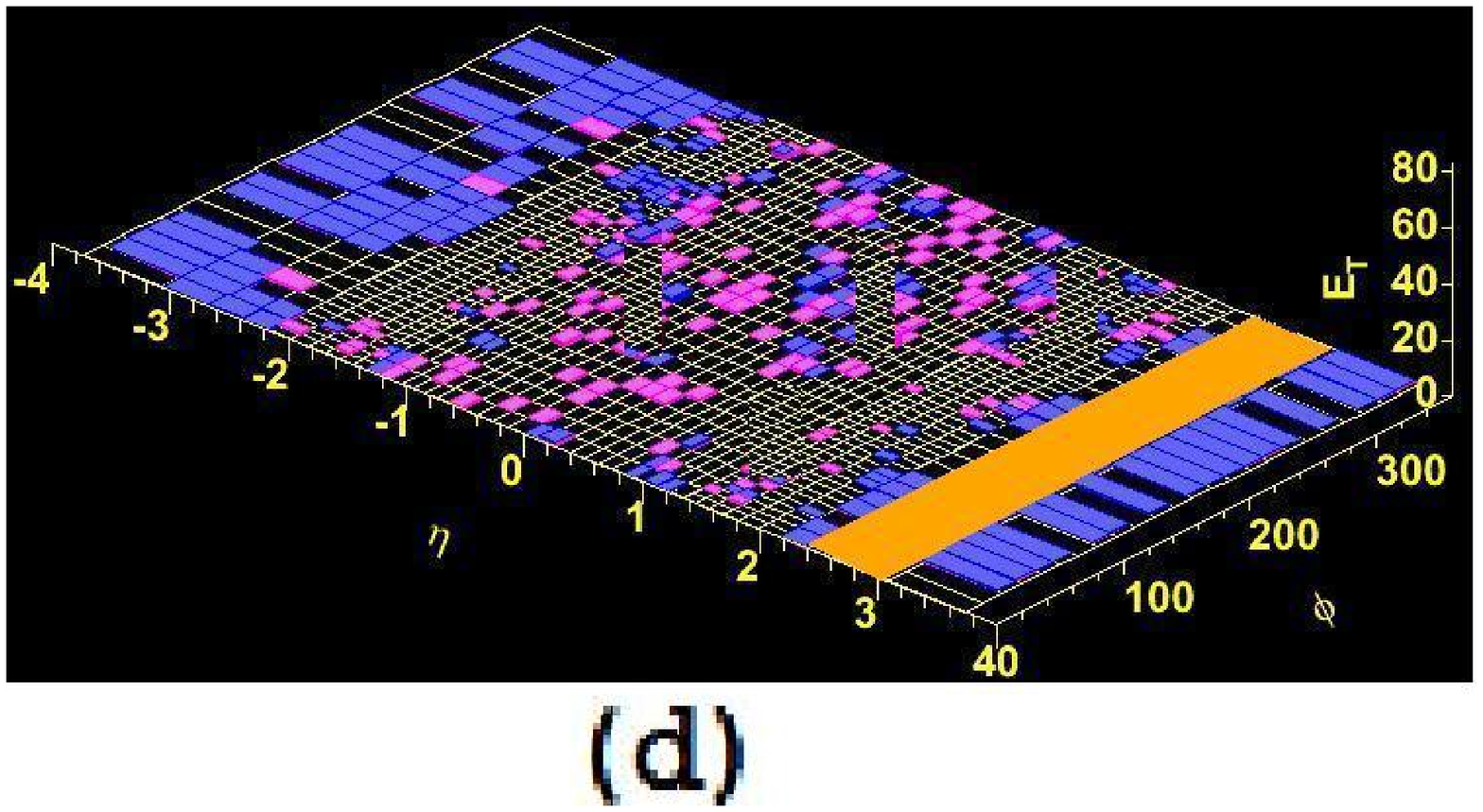}
\caption{(a) $p_{\perp} > 200$ MeV, (b) $E_T > 500$ MeV,
 (c) $E_T > 100$ MeV (d) $Z^0Z^0$
production region
 \label{fig4}}
\end{figure}

The first CDF $Z^0Z^0$ event, shown in Fig.~4, was recorded in 2004 - before pile-up!!!
It was first counted as a $Z^0Z^0$ event, then rejected because one electron was insufficiently isolated,
and finally counted. It has some remarkable features. A cut-off
$E_T > 500$ MeV  leaves only a few extra particles, but with
$E_T > 100$ MeV , many more ($>$ 70) fill the rapidity axis
away from the very forward $Z^0Z^0$ production region which, as shown in Fig.~4(d), is almost out of the
detector. While this is {\it just what would be expected for QUD events}, a 4 electron event would have been very rare in the Standard Model, with the very small accumulated luminosity. Also the very high hadron multiplicity (which was discovered serendipitously) is unexpected in the Standard Model. Unfortunately, pile-up made looking for similar events at the Tevatron impossible. So, was this event part of a (QUD predicted) forward x-section that was almost
entirely missed?  

The first CMS $Z^0Z^0$ event (4
$\mu's$), shown in Fig.~5, was recorded when the accumulated luminosity was
$\sim$2-3 pb$^{-1}$. Therefore, naively, from $\sim$25 fb$^{-1}$ we might expect $\sim$
10,000 $Z^0Z^0$ events, yet only $\sim$ 400 have been seen!!
\begin{figure}
\centering
\includegraphics[width=0.4\columnwidth]{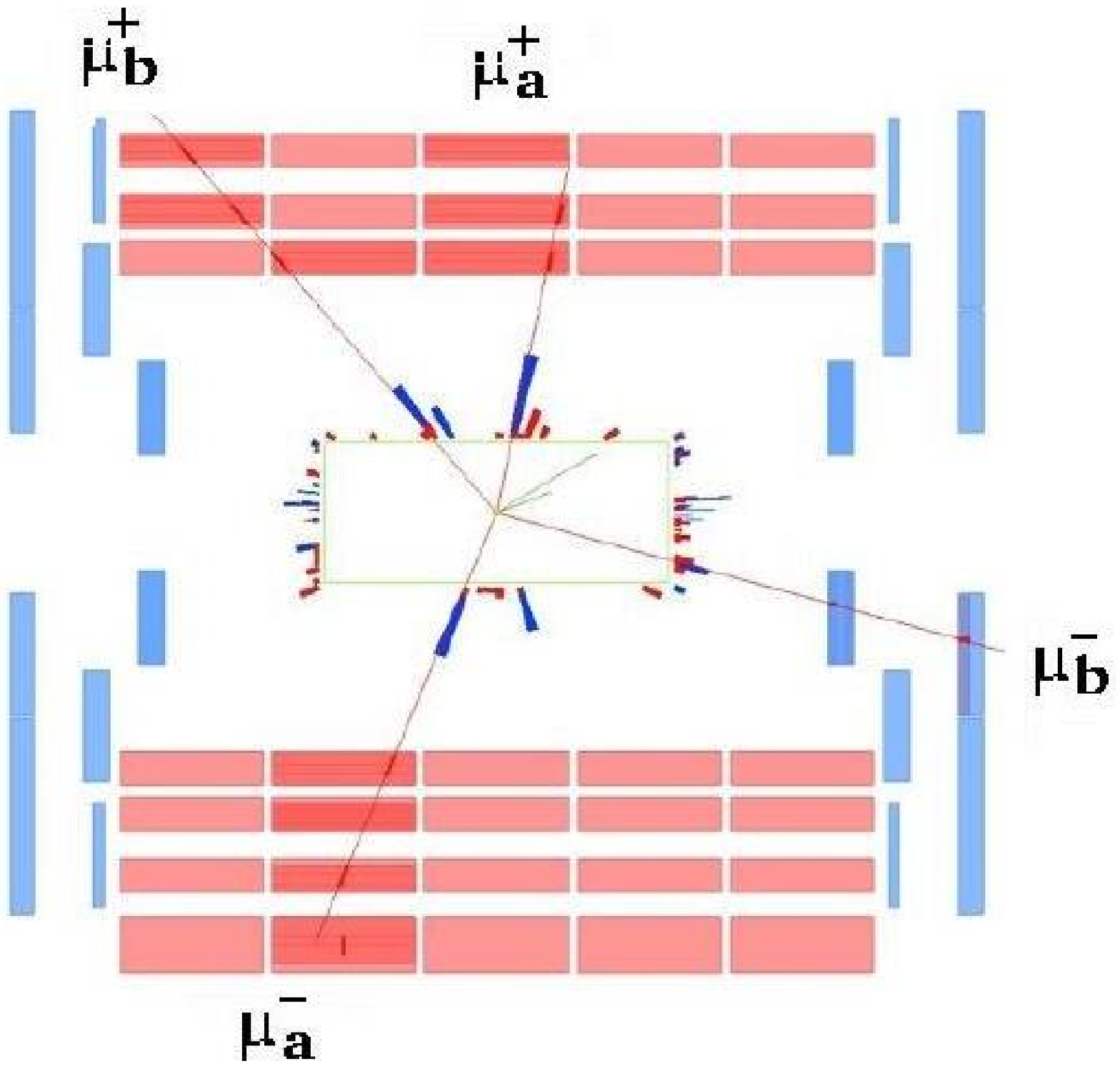} 
\includegraphics[width=0.4\columnwidth]{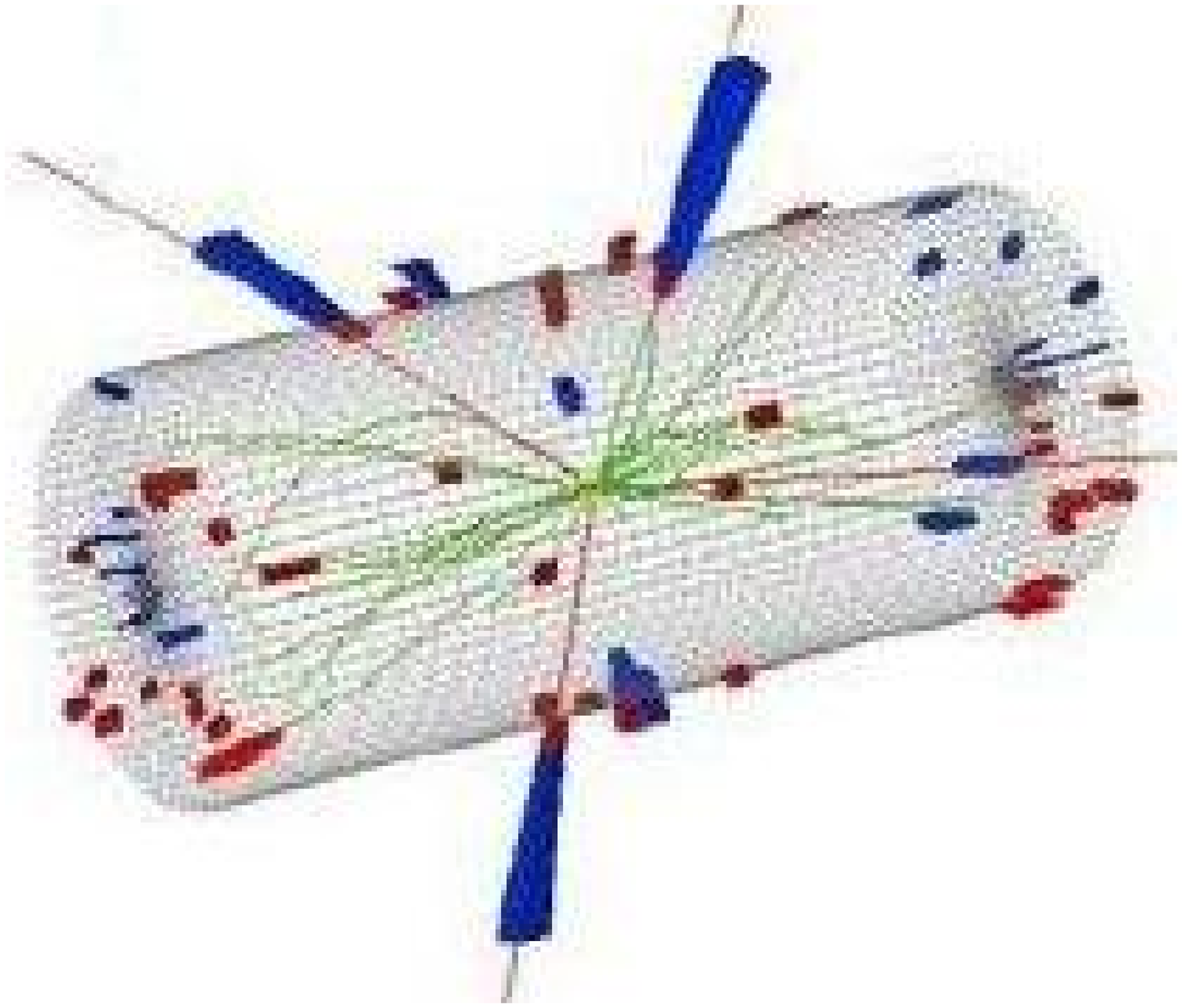}
\caption{The First CMS $Z^0Z^0$ Event \label{fig5}}
\end{figure}
It is a remarkably clean event With $p_{\perp} > 1$ GeV there is only 
two extra particles. With no cut-off there are twenty additional particles which all have momenta in one, or the other, of the two forward directions. Also $<n> and <p_{\perp}>$ are close to minimum bias. Moreover, both $Z^0$'s are very central and $p_{\perp}$(ZZ) is unusually low $\sim$ 3 GeV.
Clearly, this does not look like a Standard Model hard scattering event! Could it
also have been part of a QUD 
x-section, containing $Z^0$ pair events distributed over a wide range of rapidities, that were 
largely unseen because of pile-up?
 
A CMS 4e event with pile-up is shown in Fig.~6.
\begin{figure}
\centering
\includegraphics[width=0.35\columnwidth]{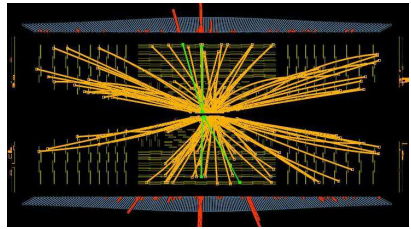} 
\caption{A CMS $Z^0Z^0$ Event With Pile-Up \label{fig6}}
\end{figure}
The line of scattering vertices is clear. Not only is it obviously impossible to determine any properties of associated 
soft hadrons produced with the $Z^0$ pair, also more forward-going leptons and photons will surely
be very difficult to isolate!

In the recent CMS-TOTEM special run, the rapidity coverage was   
\begin{center} 
{\small CMS: $|\eta| < 5.5$,
T1: $ 3.1 < |\eta| < 4.7$, 
T2: $5.3 < |\eta| < 6.5$,
FSC: $6 < |\eta | < 8 $}
\end{center}
For events with clearly isolated forward-going protons, and with rapidity gaps imposed outside of T2, the central $M_{\scriptstyle TOT}$ calculated from TOTEM Roman pot measurements 
was compared with $M_{\scriptstyle  CMS}$ measured in
the central CMS detector.  
In events where $ M_{\scriptstyle CMS} << M_{\scriptstyle TOT}$
corresponding additional tracks were generally seen in
the TOTEM T2 detector. However, in some events no
additional tracks were seen and in a few
the missing mass was as high as O(400) GeV. Could these events include
$Z^0Z^0 \longrightarrow 4 \nu$'s, as part of the large rapidity QUD  $Z^0Z^0$ x-section? 

\section{The Low Luminosity Future?} 

QUD x-sections should increase with energy, but 
increased high luminosity could still hide signals and, moreover, low luminosity runs will be short and focus  on small $p_{\perp}$ physics. However, CMS-TOTEM is working well - with beautiful double-pomeron multi-jet event displays
and also ``missing mass'' events recorded. Assuming a major part of the x-section has been missed at high luminosity, as I have argued, then some $Z^0Z^0$ and $W^+W^-$ pairs could be seen in the CMS detector, even at low luminosity.
The unprecedented wide rapidity coverage
of rapidity gaps and hadron multiplicities suggests that direct evidence for the link between pomeron and electroweak physics, that I have described could be seen! 

If the ``nightmare scenario'' persists after extensive high luminosity running, and
significant evidence of new phenomena is seen in brief low luminosity runs, we might hope (very optimistically) that, eventually, there could be a transition to 
full-time low luminosity - with modified detectors?? If it is present, QUD physics would provide a rich and exciting program with a wide variety of
phenomena.
 
\section{Some QUD Virtues} 

{\openup-0.5\jot
\begin{itemize} \openup-0.25\jot{
\item{\bf QUD is self-contained and is either entirely
right,
\newline 
or simply wrong!}
\item{\bf The scientific and aesthetic importance of an 
\newline 
underlying
massless field theory for the Standard Model 
\newline 
can not be exaggerated.}}
\end{itemize}}

If hard evidence of an electroweak scale strong interaction
appears, supporting the existence of QUD, 
there will be a {\it (perhaps needed?)}
radical redirection of the field, from the pursuit of rare, elusive, probably non-existent, short-distance physics, to the full-scale study of novel high-energy, unexpectedly large x-section, long-distance physics.
 
Assuming the QUD S-Matrix can be derived as I have outlined, then -
 \begin{enumerate} {\openup-1\jot
\item{\it The only new physics is a high mass sector of the strong
interaction that gives electroweak symmetry breaking and dark matter.}
\item{\it Parity properties of the strong,
electromagnetic, and weak interactions are naturally explained.}
\item{\it The massless photon partners the ``massless'' Critical Pomeron.}
\item{\it Anomaly vertices mix the reggeon states. Color
factors could produce the wide range of Standard Model
scales and masses,with small Majorana neutrino masses due
to the very small QUD coupling.}
\item{\it Particles and fields are truly distinct. Physical
hadrons and leptons have equal status.}
\item{\it Symmetries and masses are S-Matrix
properties. There are no off-shell amplitudes and there is no Higgs field.}
\item{\it As a massless, asymptotically free,  
fixed-point theory, QUD induces
Einstein gravity with zero cosmological constant.}}
\end{enumerate}




\begin{thebibliography}{99}


\bibitem{arw} A.~R.~White, Considerable background, including references, is provided in a series of papers, arXiv:1106.5662; arXiv:1009.4850; arXiv:0803.1151; arXiv:0708.1306; arXiv:1206.0192; arXiv:1301.5628.

\end{thebibliography}
\end{document}